# KINEMATICS GROUNDED ON LIGHT

Z. Néda

Babeş-Bolyai University, Department of Physics, Cluj-Napoca, Romania
zneda@phys.ubbcluj.ro

**ABSTRACT**

*The space-time of modern physics is tailored on light. We rigorously construct the basic entities needed by kinematics: geometry of the physical space and time, using as tool electromagnetic waves, and particularly light-rays. After such a mathematically orthodox construction, the special theory of relativity will result naturally. One will clearly understand and easily accept all those puzzling consequences that makes presently the special theory of relativity hard to digest. Such an approach is extremely rewarding in teaching the main ideas of Einstein's relativity theory for high-school and/or university students. Interesting speculations regarding the fundaments and future of physics are made.*

## ON PHYSICS AND POSTULATES

Physics is undoubtedly an experimental science [1]. It's laws are derived from experiments and it's theories are confirmed through experiments. It is considered to be exact science, but many scientist consider it less rigorous than mathematics is. The main reason behind this fact is that mathematics is built on clear postulate systems, while in physics it is not always obvious what our postulates are. This is also the way how we teach physics: it is not straightforward for students (and sometimes also for the teacher) what we postulate and what results from these postulates. From time to time we use phrases like "it is evident", "it is not unrealistic to assume", or simply "let us assume". In such cases students might get confused whether we state a new postulate or affirm something realy obvious resulting logically from the already accepted postulates.

Any logically consistent theory must be based on postulates. It is not possible to prove everything, and there is thus no ultimate true [2]. Any statement or law in science is true or false in the framework of a postulate system. In physics we also have thus postulates. Postulates can change as physics progress, and our goal to reduce the number of postulates and/or to postulate more simple things is slowly, but advancing. The physical description of nature is however always bonded to a postulate framework. We might or might not realize these postulates, but they are silently there.

The less rigourous manner in which physics handles it's postulates was however not an impediment for its development. Being non-rigorous in the pure mathematical sense has also advantages and allows for describing more easily the complexity of nature using just our intuition. This is one reason why physics became so successful and advanced quickly in describing many complex phenomena. Not realizing however the postulates on which our reasoning in physics is built and relaying to much on intuition can seriously fool us and obvious facts might appear as paradoxes. There are times thus in physics when we have to stop, look back and review the postulate system on which our current understanding is built. In this work I plan to do this, by building the basic elements of modern kinematics in a rigorous and logically consistent manner. I will argue that the tool we use nowadays to define the geometry of the physical space and time is light, or electromagnetic waves in general. After taking the effort of doing a solid grounding by postulates for our main tools: space and time, we then collect the benefits. The special theory of relativity, Lorentz transformations and all "strange" consequences of these will become crystal clear. We will be surprised how simple and logical and self-consistent this theory is. We will also realize some common misconceptions that are still sorounding us in many textbooks and science popularization books or movies.

**LIGHT AND ITS' SPEED OF PROPAGATION**

Light and electromagnetic waves in general are our primary source for getting information about the surrounding world [3]. Light is however a tricky entity. Light rays propagate in straight line, and it shows both wave and particle properties. In some experiments light behaves as a classical wave (diffraction, interference) and yet in other experiments exhibits particle properties (photoelectric effect, Compton scattering). There are also properties of light that can be understood both by it's particle and it's wave-like nature (refraction, reflection). Nowadays we know that light is an electromagnetic wave, which transports energy in quanta, named as photons. Electromagnetic waves and therefore the propagation of light are described by the celebrated Maxwell's equations [4] which predicts that the propagation speed of electromagnetic waves in vaccum, and light in particular, is very big (~300 000 km/s). Experimentally it was quite a hassle to determine its value and a lot of debates were on this subject during the history of science (see Appendix I.).

Getting the value for the speed of light is however only half way the problem done! Whenever we speak about speed we do need to specify the reference frame relative to which this is measured. Sound and acoustic waves propagate relative to a medium with mechano-elastic properties. In air for example, when we admit that the speed of sound is 340m/s this is understood as being the value measured relative to air. We are used to the fact, that each wave-like phenomenon propagate relative to some medium, and when we speak about the velocity of the wave, we usually do refer to its velocity relative to this medium. We know now that light is an electromagnetic wave, and the natural question we can pose is to find the medium in which electromagnetic waves are propagating, or the medium in which Maxwell equations are valid. The medium was hypothetically named *aether* and the hunt for its discovery began. The method to reveal the existence of eather was based on the detection of the eather wind. The effect of the eather wind on the propagation of light would be something similar like the effect of the wind on the propagation of sound. For sound waves in air we know that an observer that travels with a speed of 10 m/s relative to the air, would measure a speed of 350 m/s for the sound waves coming in his/her direction and a speed of 330 m/s for the sound waves travelling in the same direction as his/her motion relative to air.

Carefully designed experiments were however unable to detect such eather wind [5]. Eather proved to be either inexistent or hard to be observable due to some unknown reasons. As we will detail in the following, physics got in an unpleasant situation, which drove physicist to rethink the very basic foundation of our basic notions in physics: space and time and the postulates on which the modern apparatus of physics is build.

**SPACE AND TIME IN PHYSICS**

Space is the most fundamental entity of physics, the ground and background of matter, our models' natural environment [6]. Its existence can be stated as a postulate: it is something that exists. Physical space does not have however an inbuilt, intrinsic geometry. Geometry is our attempt to describe the form of the objects and their relative positions in space, following the idealisation we have achieved in mathematics. Physics relies on geometry, since one of our aims is to account for the relative position of objects and to describe their changes. In physics we have to distinguish clearly between the mathematically defined abstract spaces and the space we deal with, when the description of the spatiality of the matter is our focus.

The **ideal mathematical space** is defined by postulating all its *metrical* properties. Metrical, in this case means that we have a distance unit and we can construct a 'geodesic' line between any two points. This geodesic line is the shortest distance between two points in our ideal space. It is the generalisation of the concept of 'straight line' encountered in simple Euclidean geometry. Once we have the geodesic lines and postulated the distance unit, we are able to measure distances between any two points. The distance between two points will be obtained by measuring the length of the geodesic line drawn between them. We also have the possibility of defining objects with spatial extension in the space: spheres, cubes, etc. Objects placed in such a space will not alter the postulated geometry.

For example, Euclidean geometry in plane defines its metric by the five postulates we all learned in elementary or high school [7]. Euclidean space is the simplest mathematical space. Geodesic lines are the "classical straight lines" we all have imprinted in our minds since our middle school years. For any triangle we have that the sum of its interior angles is 180 degree. In the Euclidean space one can also state that for any elementary triangle built on two infinitesimally small segments perpendicular on each other: *dx* and *dy*, the third segment (*ds*) can be calculated as: $ds^2 = dx^2 + dy^2$.

Our **physical space** is not an idealized mathematical space, and metrics is not an a priori given; rather, it is determined by the matter spread throughout the Universe. In order to construct any geometry in this space, the first task is to reveal the geodesic lines and have a well defined concept of distance. One possibility would of course be to postulate an a priori metrics for space, and the handiest would naturally be a three dimensional Euclidean space. In such an approach distances between objects would be measured using a hypothetical infinitely extensible ruler (which is postulated to be the straight line) and a postulated unit of length. This is what Newtonian physics does. Distances between the bodies around us are measured using such an ideal ruler together with a postulated unit of length, which is the International Prototype Metre kept at the International Bureau of Weights and Measures in Paris. Apart from the logical inconsistency of the a priori postulated metrics (straight line concept), the drawback of this approach is that it makes impossible to measure distances on a cosmological or micro scale. Using a ruler you cannot measure large distances on Earth, and you definitely cannot extend investigation to the cosmological or microscopic scale. A viable alternative to this approach is to use light rays to reveal the geodesic lines in the physical space. Light rays in vacuum have always been thought to move in a 'straight line', or at least this was considered to be true on the terrestrial scale in an optically homogeneous medium such as air or a vacuum. After the development of Euclidean geometry, measurement of large distances and methods of mapmaking, such as the classical triangulation method, employed this idea extensively. The solution seems simple. Then, why not turn the problem around, and postulate the path of light rays between any two given points as a geodesic line? The idea could work on both the cosmological, terrestrial and micro-scale and would define a usable geometry for physics.

Nowadays, we need to be careful when speaking about the geometry of physical space, since we inevitably mix up two different metrics... *In our everyday life* we use the Newtonian paradigm of ideal space alongside Euclidean metrics, which we find convenient. The optically different materials around us and, as a result, the light refractions at the boundaries of different media make complicated the use of geodesic lines defined by light. This is why we tend to say that light rays bend and we accept that sometimes they do not move in a 'straight line'. Practical problems relating to tracking the path of light rays and handling such rays also suggest that it is easier to use a traditional ruler. But we should add here, that given the rapid advances in laser technology this picture is quickly changing; even on a small scale we nowadays use light to measure distances and trace straight lines. *On a cosmological or microscopic scale*, however, the concept of the straight line is inextricably linked to the path of a light ray. The metric is defined by the propagation path of light, which is influenced by the matter spread throughout the Universe. Luckily, in our everyday life the two different metrics are not at odds, since light rays do not bend in any detectable way in a homogeneous medium and the gravitational field around us is rather homogeneous over small distances. However, if light interacted more strongly with the Earth's gravitational field and this gravitational field were not homogeneous, the situation would have been more complicated. We would immediately sense there was problem, since our ruler would no longer look 'straight' to us, even if the medium around us were a vacuum.

Light therefore defines the geometry of the physical space. Such geometry may not be simple, since the metrics is influenced by the interaction between matter and light. The properties of light are inevitably mapped onto the geometry of space. We might well discover that in such space the sum of the interior angle of the triangles is not 180 degree and for the diagonal *ds* of the elementary triangle constructed on the two perpendicular segments *dx* and *dy* we find that $ds^2 \neq dx^2 + dy^2$. Instead of this relation, we might find in a plane a more complicated formula of the type: $ds^2 = a(x,y) \cdot dx^2 + b(x,y) \cdot dy^2 + c(x,y) \cdot dx \cdot dy$. In such cases the geometry of the physical space is a non-Euclidean type. The abstract mathematics of non-Euclidean ideal spaces developed by Bolyai, Gauss, Lobachevski and Riemann therefore acquired a major application in dealing with the geometry of the physical spaces [8].

**Time in the physical space.** In order to approach the concept of time we should first think about what gives physical meaning to it. We use the concept of time, in order to describe changes [5]. In a completely static Universe, time would have no role. More precisely, we need time so that we can order the events or changes that happen in the physical space, to detect and describe their possible *causality*. The existence of causality is a basic postulate in science. Causality implies that the cause must always precede the effect in time, and the same causes always produce the same effects. Time is a creation of our mind and makes sense only in connection with changes and movement. It is derived from two basic quantities that do exist in the Universe: space and changes. Time is therefore not an a-priori given quantity and similarly with the geometry of the physical space, the Universe does not naturally come with time embedded in it, and it is something that we have to construct it.

*Clocks* are devices that are used to measure time [9]. In order to make time measurable we need an etalon clock that gives periodic signals. Once we have defined this etalon clock, the unit of time will have been properly fixed, and we will be able to time events taking place in the physical space. However, an etalon clock is just halfway in timing events. Physical events take place at different points in space and in order to describe them properly, we need synchronised 'clones' of this ideal etalon clock placed at each point in physical space, so that we can time events at their own spatial location. Synchronisation of clocks and thus measurement of the local time of events was—and still is—a difficult task. Let us merely scratch the surface of the problem's complexity. In order to have reliable local clocks, ideally you first have to produce clones of the original master clock, and than from time to time you have to synchronise them with the master clock. In order to achieve synchronisation an ideal information carrier is needed. For this information carrier you need to know exactly the speed of propagation along the geodesic line between any two clocks, since only in this way can you account for the time lapse in the propagation of the information. We therefore also have to locate the geodesic line between the clocks and calculate precisely the distance along this line. All these are of course needed if we have an information carrier that propagates at finite speed. On the other hand however, if we had an ideal information carrier that propagated at infinite speed, the problem would be much simpler, and synchronisation of local clocks using such a signal would be trivial. Alternatively, timing of events occurring at different points in space can be hypothetically achieved using only one master clock. To proceed in this way, we again need some universally available information carrier and we have to know the length of the geodesic line between the master clock and the event. When the event occurs, we transmit a signal through the information carrier to the master clock. When the master clock receives the signal, we record that time and subtract from it the time needed to travel the distance between the event and clock. This procedure requires also the knowledge of the geodesic line and the speed at which the signal is propagated along it. Timing is again simple if we use an information carrier that travels at infinite speed.

It was only natural that in the beginning physics considered the simplest alternative: definition of an ideal time. This was defined at each point in space by hypothetically similar and ideal clocks, based on the assumption that they were synchronised with an ideal information carrier that travelled at infinite speed. Alternatively, this was equivalent to time events with a master clock, assuming no time lapse between the occurrence of any event and its detection at the location of the master clock. The simplicity of Newtonian mechanics resides in the fact that we assume the existence of an ideal information carrier that moves at infinite speed. Unfortunately, however, our Universe is more complicated than that. The fastest and the only universally available information carrier we are nowadays aware of is the electromagnetic wave, and we do know that its propagation speed is finite. In such case there is a serious catch, however: we do not know the medium through which light is propagated, since we were unable to detect aether. Whether it exists and we are just unable to find it or whether it does not exist, there is still a problem. In order to be capable of timing events at different spatial locations and to have a well-defined space time in physics we have to know the propagation speed of light along the correct geodesic line in any reference frame. The aether wind would influence this propagation speed, and the timing of events would then depend upon the frame of reference where the master clock is positioned. Now imagine for a moment that we know where the aether is. Theoretically, we would then be able to make a correct timing, but in practice it would be an arduous task. However, the advantage would be that by this method we would have an absolute frame of reference related to aether, and the timing of events relative to this would preserve a time concept that would also be absolute in nature. At the beginning of the twentieth century, however, aether was undetectable, and in order to progress, physics keenly needed a properly defined concept of time in physical space, if it was to tackle cosmological and electromagnetic phenomena.

Einstein gave in 1905 the only feasible outcome from this paradoxical situation by postulating that the speed of light in vacuum is a universal constant independent of the direction of propagation or frame of reference [10]. In the context of how we constructed and introduced the logic of time, this step now seems quite obvious and it is easy to conclude that physics had no other alternative. Thitherto space-time had not been properly defined, it did not make sense to speak about speed, and so the speed of light too had no physical meaning without a properly defined time. Einstein's postulate, as well as fixing the value of the speed of light, allowed local times to be properly defined at each point in physical space and the "catch 22" type paradox was elegantly solved. In the beginning, the value of the speed of light was set at the best available experimental data, and it was proposed that this value should be constantly revised according to new experiments.

The story of grounding the physical space-time by means of light rays does not end here, however. For a consistent definition of space-time one immediately realises that it is problematic that geodesic lines and time are defined by means of light propagation, whereas the unit of length is still defined by the traditional ruler. In addition, if we know exactly the speed of light (since in Einstein's paradigm it is a universal constant), then the unit of length should result automatically after the time unit has been set. The time unit of the master clock has to be set in any case, both in the Newtonian and in the Einsteinian paradigm in order to have a working master clock. By setting the time unit [11], and by knowing exactly the speed of light we could measure any distance in space using light rays and an etalon clock. Instead of postulating the unit length, it is much more convenient for modern physics to postulate the value of the speed of light. It was not until 1983 that physics, accepting the proposal of the Hungarian physicist Zoltan Bay, decided to postulate this value in consensus with the best experimental results at that time [12]. Today this value is set at $c=299,792,458$ m/s. This was the last crucial step in defining a logically consistent space-time for physics.

However, this last step closes the circle and makes it impossible henceforth to detect aether by the eather wind, should it exist. In our new measurement paradigm, both the physical definition of distances and the construction of space-time are linked to the fact that light travels at a constant speed in every direction through space, independently of the chosen frames of reference. Any self-consistent eather wind experiment should give thus negative results. We do have thus a usable and well-defined space-time entity, but we have lost the aether-wind in the presently postulated physical space-time.

**THE PRINCIPLE OF RELATIVITY**
In physics we may use several frames of reference for describing the same phenomenon. For example, the collision of two cars on a road can be described within the frame of reference of the road, within the frames of reference of the two colliding cars, or within the frame of reference of any other car driving past on the road. It can be even described within the frame of reference of a light ray passing at the speed of light, or infinitely many other imaginable reference frames. These frames of reference are in motion relative to each other. When describing events from different frames of reference we intend first to locate the event within the particular frame of reference (giving the spatial and time-like coordinates) and then we try to understand it in a causal way according to previous events, applying the laws of physics. Each frame of reference comes with it's own space-time entity, so we are interested to get also the relations (transformations) that connect the spatial or time-like coordinates of any event as measured from different reference frames.

Mechanics is founded on the experimental observation, which holds that there are special frames of reference within which the law of inertia is valid. These reference frames are called *inertial frames of reference* and these are those special reference frames where the geometry is Euclidean. If one frame of reference is inertial, any other frame of reference that is moving at a constant speed relative to it is again an inertial frame of reference (according to our postulates defining the geodesic lines their geometry is also Euclidean).

For the inertial frames of reference physics accepted the *principle of relativity*. According to this principle all inertial frames of reference are equivalent when describing physical phenomena, meaning that the laws of physics are the same in all inertial frames of reference. Mathematically this implies that the basic equations describing physical phenomena should have the same form in all of them. Taking into account their Euclidean nature (similar metric properties) this seems to be a logically consistent statement. According to the principle of relativity there is also no distinguished inertial frame of reference, or in other words there is no *absolute frame of reference*. Motion is always relative, and whenever we speak about motion we mean it only in a relative manner. Contrariwise, if we had an absolute frame of reference the notion of motion and rest would also be absolute. In Newtonian space-time the absence of an absolute frame of reference is due to the fact that the information carrier travels at the same infinite speed relative to all inertial frames of reference. In Einsteinian space-time the absence of an absolute frame of reference is a direct consequence of the fact that we have postulated the constancy of speed of light independently of the frames of reference. In such a way, we postulated that there is no aether, which would have been the absolute frame of reference for physics. Our postulate about the uniform propagation speed of light, independently of the frames of reference, is thus consistent with the principle of relativity, which holds that there is no absolute frame of reference in physics.

**COORDINATE TRANSFORMATIONS**
Both Newton's ideal space-time and Einstein's tangible space-time ought to be consistent with the principle of relativity. In both cases the spatiality and temporality of events are completely measurable and we can construct a well-defined system of coordinates to characterize this. However, the principle of relativity does not affirm that the temporal or spatial coordinates of events measured from different inertial frames of reference are the same. In Newtonian space-time, all the local clocks of the inertial frame of reference are naturally synchronised and time is absolute, but spatial coordinates are not. In Einstein's Universe the concept of the synchronisation of clocks in different inertial frames of reference is more complicated and consequently time is not absolute. In both cases, however, if we know the spatial and temporal coordinates in one inertial frame of reference, we can calculate these coordinates in any other inertial frame of reference that is in motion relative to the original one at a given velocity. We call this *coordinate transformation* between two frames of reference [13] (for deriving the coordinate-transformations see Appendix II).

For the Newtonian space-time the coordinate transformations are given by the simple Galilean transformations we all learned in high school. Within the framework of Galilean transformations time coordinates are invariable. The coordinate transformations for Einsteinian space-time are given by the more complicated *Lorentz transformations*. In this case, time coordinates are not invariable, and they are transformed as a whole along with spatial coordinates. This is why we say space and time forms a 'space-time continuum'. The consistency of the space-time defined by Einstein is demonstrated by the fact that the basic equations of electrodynamics, the Maxwell's equations [4], are fully covariant with Lorentz transformations. From this viewpoint, the electromagnetic theory of light becomes a logically consistent and closed system. Lorentz transformations were derived by assuming there is no absolute medium through which light is propagated (no aether), and the equations that describe light lead us to the same conclusion: within each reference frame light travels at the same universal speed.

**THE AFTERMATH**

If one uses the space-time entity tailored on the finite propagation speed of light (and implicitly the Lorentz coordinate transformations), several fascinating consequences arises. First, one can easily realize that the simultaneity of events is not an absolute concept anymore. Events that are simultaneous in one frame of reference will not necessarily be simultaneous as viewed from other reference frames. The time-length of an event, or the spatial length of an object becomes also a relative concept. An observer moving with respect of a purely time-like event (an event happening at the same spatial coordinate in a reference frame that is in rest relative to the event) will measure a longer time-length for the event than the observer in rest relative to that event, a phenomenon called *time-dilation*. The spatial length of an object measured by an observer in rest relative to that object is longer than the one measured by an observer moving in the direction of the measured length. This phenomenon is known as *length-contraction* (Appendix II). Composition of velocities is also more complicated than in the Galilei-Newtonian kinematics. The interested reader can learn more about these fascinating effects from many excellent textbooks (see for example Einstein's original text [13]), here we will not elaborate more on them. All these effects that seemingly contradicts our common sense are simply the direct consequences of the method by which our actual space-time entity is tailored on light.

As we have argued in the previous chapters, the story and the logic of the construction for the physical space-time is quite complicated, and even physicists sometimes forget this. Once it has been properly understood, however, special relativity becomes crystal clear and one can also realize that the presently used space-time paradigm might not be the ultimate solution for grounding physics. In the followings I will allow myself to be less rigorous and finish with some speculative thoughts.

An important, and sometimes misunderstood issue that we have to discuss here is the existence of a signal other than electromagnetic waves for allowing information to travel on cosmological scales of length. As our detection methods improve thanks to rapid advances in technology, a channel via such a non-conventional information carrier might become more appropriate for communication and measurement. At present, nothing guarantees that such an information carrier, if it exists, would travel at a lower speed than light. Contrary with what it is falsely believed, a speed bigger than the speed of light wouldn't be a disaster for modern physics. The problem that might arise in such a case, would be that the space-time we built in the context of the theory of special relativity would not be self-consistent and we would violate the principle of relativity. It would be inertial reference frames in which one would detect the cause preceding the effect, contradicting the principle of causality (see Appendix III). In order to make the theory consistent, instead of light we would now use the new information transmitter signal to synchronise clocks. Postulates regarding this new signal would replace those made for light. The advantages from such a new space-time paradigm would be that we would be able to answer some of the questions we have previously labelled 'unanswerable'. Detection of the aether wind, the universal nature of the speed of light, and straight-line propagation of light rays would become questions that once more made sense.

Escaping the bonds of Einstein's space-time and using some other logically consistent definition of space-time is an interesting challenge even if we cannot find any new information carrier. A new method of synchronising clocks or a new procedure for measuring time, along with the traditional ruler for measuring distance, might also help to bring the aether wind to light. Such experiments would not use interference or other wave-like properties implicitly based on Lorentz covariant electrodynamics. Instead, direct time of flight measurements of the propagation of light-rays, performed on laboratory scales of length, would be the most appropriate for us. As optoelectronics is making rapid progress, this does not seem to be unrealisable in the near future.

Previously we argued that our physics is based on the presumption that we have no absolute frame of reference, and all inertial reference frames are equivalent regarding the laws of physics. This also means that there is no special frame of reference known to us, relative to which electromagnetic waves are propagated. Modern cosmology is cautious enough to teach us that this might not be true, however. Apparently, there is a special, absolute frame of reference, relative to which the Cosmic Microwave Background Radiation (CMBR) is isotropic, and this could be regarded as an absolute frame of reference for the electromagnetic waves originating from the birth of the Universe [14] (more information on CMBR is given in Appendix IV). The special frame of reference defined by the CMBR naturally raises several questions and thoughts about the present foundation of physics. Firstly, it would be important to know whether the CMBR's frame of reference generally defines the aether we were looking for. Will light rays travelling through the Universe always move with constant speed only relative to this frame of reference? When investigating such questions, we have to be quite cautious, however, since our present working paradigm, built on the constancy of the speed of light, would inevitably distort the measurements and their interpretations. Nonetheless, the proven existence of a special frame of reference for the CMBR is a serious argument pointing to the need to overhaul the present foundation of our basic concepts of space and time.

Another important issue we might discuss here in connection with space-time based on light is the microcosm. At the level of elementary particles, quantum mechanics seems to work when describing the microcosm. The image currently provided by physics, which states that elementary particles are wave-like in character and at the same time behave like classic particles is merely a desperate attempt to create a visual picture. Let us make a less desperate attempt and provide an alternative understanding of elementary particles' strange dual nature starting from the obvious fact that the tools for investigating the microcosm are also electromagnetic waves. All physical information is collected via this channel, thus the geometry of the microcosm is shaped by light. Distances are measured via electromagnetic interactions and electromagnetic waves. We do know, however, that light has an experimentally proven dual nature. Apart from being a particle, its propagation is also wave-like. On the microscopic scale (the scale of length comparable with wavelength of light) this wave-like behaviour dominates, with the result that photons (energy carriers of the light) can be propagated in a strange manner along non-deterministically defined paths between two points of the space. The straight-line concept based on light yields a fuzzy, non-deterministic trajectory. If we do extrapolate our accepted definition of geodesic lines on microscopic scales, this thinking leads us to a non-deterministic, random metrics on the microscopic scale. The movement of microscopic objects would automatically obtain a probabilistic character, something similar to what we see in quantum mechanics. From this viewpoint, geometric spaces with random metrics might be than the tool to provide the particle-wave duality with an alternative description.

**APPENDIX I. MEASUREMENT OF THE SPEED OF LIGHT**

Ancient Greek philosophers were much more interested about vision, and light was only their second concern. Empedocles for example presumed that light travels with finite speed, while Aristotle assumed an infinite propagation speed for light. The first scientifically grounded thinking about the finite light propagation speed was formulated by the Arab scientist Ibn al-Haytham in the eleven century, and the first documented attempt to measure its value was by Galileo Galilei around the year 1600. Galileo designed a very basic experiment with too lamps placed on the top of two hills sufficiently far away from each other, but still in sight. His method was the following: let in the beginning both lamps be covered. Then, he would uncover his lamp, having instructed his apprentice to uncover his lamp as soon as the light from Galileo's lamp become visible. By this experiment Galileo planned to measure the time-gap between uncovering his lamp and the receiving signal from the other lamp. Due to the extremely high propagation speed of light, this method of course couldn't work, even if the two hills are tens of kilometers apart from each other. Galileo finally concluded that if the speed of light is finite, then it is too high to be measurable in such experiments. He also concluded that the speed of light must be at least 10 times bigger than the speed of sound, and we know nowadays that this is a strongly underestimated limit. Danish astronomer Olaf Roemer gave the first scientific proof that light travels with a finite speed. He made this conclusion by studying the observed eclipse times of the Jupiter's moon:

Io. Astonishingly based on his results a good estimate can be obtained on the propagation speed of light. The first successful experiment for measuring the propagation speed of light in terrestrial experiments was made by the French physicist Armand Fizau in the mid-nineteenth century, obtaining a result that differs with less than 1% from the value accepted nowadays. Many later, carefully designed experiments confirmed these estimates and improved its accuracy. Nowadays our International System of Units is built in such manner (and we will understand later why this is a natural choice!) that the propagation speed of light is postulated. Following the proposal of the Hungarian physicist Zoltan Bay, in 1983 the propagation speed of light in vacuum was postulated as: 299 792.458 km/s.

**APPENDIX II. DERIVING THE COORDINATE TRANSFORMATIONS**

We consider two inertial reference frames, K and K'. The geometry in both reference frames is Euclidean and we construct the usual Ox, Oy and Oz coordinate axes in K and the O'x', O'y' and O'z' axes in K', so that the corresponding axes are parallel with each other. Let us assume that K is moving with a uniform speed $u$ relative to K' in the direction of the common Ox -- O'x' axis so that at time $t=0$ we synchronize the etalon clocks in K and K' ($t'=0$) and at this time moment the centre O is coinciding with O'. Let an event P happen on the Ox -- O'x' axis. Assuming that the time and length units used in both reference frames are the same, we are interested in the relation between the coordinates of the event measured in K ($x_1, y_1, z_1, t$) and K' ($x_1', y_1', z_1', t_1'$) (see Fig. 1). In our geometry for the P event $y_1 = y_1' = 0$ and $z_1 = z_1' = 0$, so we are looking for the relation between $x$ and $t$ coordinates measured in K and K'. Assuming a general functional relationship between the $(x_1, t_1)$ and $(x_1', t_1')$ coordinates we can write:

$$x_1' = F(x_1, t_1, u)$$
$$t_1' = G(x_1, t_1, u) \quad (1)$$

Using the principle of relativity, the same functional relationship should be true inversely, if we replace $u$ by $-u$:

$$x_1 = F(x_1', t_1', -u)$$
$$t_1 = G(x_1', t_1', -u) \quad (2)$$

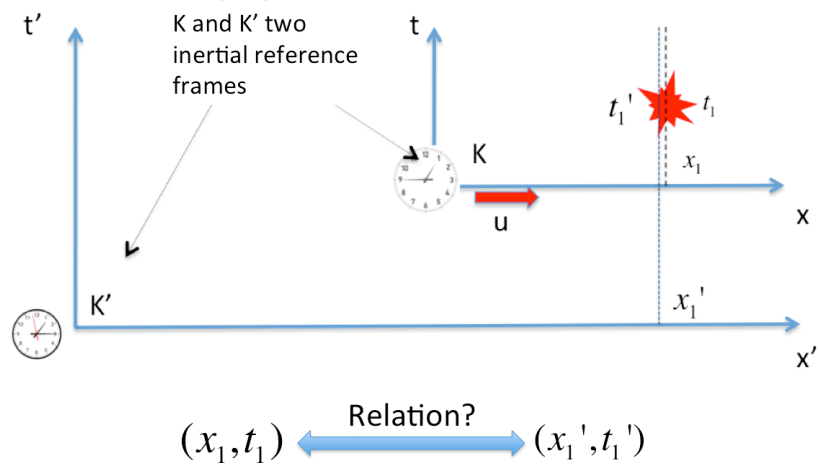

Fig.1. Coordinate transformation between the inertial reference frames K and K'.

One can get the inverse transformations however also from (1). In order that these should have the same functional form it is necessary that the transformations should be linear in the $x$ and $t$ coordinates, otherwise the functional form would change, and we would contradict equation (2) and implicitly the principle of relativity. The most general linear transformation we can write is:

$x_1' = a(u)x_1 + b(u)t_1 + d$

$t_1' = g(u)x_1 + f(u)t_1 + k$

Considering the initial condition $x_1 = x_1' = 0$ at $t_1 = t_1' = 0$ we get $d=k=0$, leading to:

$$\left. \begin{array}{l} x_1' = a(u)x_1 + b(u)t_1 \\ t_1' = g(u)x_1 + f(u)t_1 \end{array} \right\} \rightarrow \frac{dx_1'}{dt_1'} = v_1' = \frac{a(u)\frac{dx_1}{dt_1} + b(u)}{g(u)\frac{dx_1}{dt_1} + f(u)} = \frac{a(u)v_1 + b(u)}{g(u)v_1 + f(u)} \qquad (3)$$

Here $v_1$ denotes the velocity for the movement of a body as measured in K and $v_1'$ the same velocity measured from K'. According to the principle of relativity the inverse transformations should be:

$$\left. \begin{array}{l} x_1 = a(-u)x_1' + b(-u)t_1' \\ t_1 = g(-u)x_1' + f(-u)t_1' \end{array} \right\} \rightarrow \frac{dx_1}{dt_1} = v_1 \frac{a(-u)\frac{dx_1'}{dt_1'} + b(-u)}{g(-u)\frac{dx_1'}{dt_1'} + f(-u)} = \frac{a(-u)v_1' + b(-u)}{g(-u)v_1' + f(-u)} \qquad (4)$$

If we study the relative motion for the origins of the two coordinate system K and K' we get that in case $x_1 = 0 \rightarrow dx_1/dt_1 = v_1 = u$ and for $x_1' = 0 \rightarrow dx_1/dt_1 = v_1 = -u$. These conditions in (3) yields: $b(u) = u \cdot f(u)$ and $a(u) = f(u)$.

**Galilei-Newton transformations**. If one assumes that the speed of light is infinite, synchronization of clocks is trivial since there is an absolute time, and in such manner for any event one should have: $t_1' = t_1$ leading to: $g(u) = 0$ and $f(u) = 1$. The coordinate-transformations write thus as:

$x_1' = x_1 + ut_1$

$t_1' = t_1$ \qquad (5)

As an immediate consequence we have than, that for any event, it's time-length is the same from any reference frame ($dt_1' = dt_1$) and the length of any object is the same from any reference frame. The length of an object from reference frame K is measured by determining it's coordinates at the same time moment, so that $dt_1 = dt_1' = 0$. As a result we get $l_1' = dx_1' = dx_1 = l_1$. The velocity transformations result automatically from eq. (3)-(5):

$v_1' = v_1 + u$

**Lorentz-Einstein transformations.** In case the speed of light is finite, synchronization of clocks can be made only by assuming that the speed of light is the same, from any reference frame. This means that a light beam that travels along the Ox axis in K should have the same *c* finite value measured both from K and K'. Equation (3) and the relations established before yield thus: $c = \frac{f(u)(c+u)}{g(u)c + f(u)} \rightarrow g(u) = f(u)\frac{u}{c^2}$

According to this we can write:

$x_1' = f(u)(x_1 + u \cdot t_1)$

$t_1' = f(u)\left( x_1 \frac{u}{c^2} + t_1 \right)$ \qquad (6)

Deriving from here the inverse transformations one gets:

$$x_1 = \frac{x_1' + ut_1'}{f(u)\left(1 - \frac{u^2}{c^2}\right)} \quad ; \quad t_1 = \frac{x_1'\frac{u}{c^2} + t_1'}{f(u)\left(1 - \frac{u^2}{c^2}\right)} \tag{7}$$

In order to satisfy the principle of relativity (and the inverse transformations should be of the form one must have: $f(u) \cdot f(-u) = \left(1 - \frac{u^2}{c^2}\right)^{-1}$, leading in our geometry to the Lorentz transformations:

$$x_1' = \frac{x_1 + ut_1}{\sqrt{1 - \frac{u^2}{c^2}}} \quad ; \quad t_1' = \frac{x_1\frac{u}{c^2} + t_1}{\sqrt{1 - \frac{u^2}{c^2}}}. \tag{8}$$

One realizes now that quite differently form the Galilei-Newton kinematics, simultaneity of events becomes a relative concept. In case of a purely time-like event taking place in K (the spatial coordinates in K are the same, $dx_1 = 0$), we get:

$$dt_1' = \frac{dt_1}{\sqrt{1 - \frac{u^2}{c^2}}} \geq dt_1. \tag{9}$$

According to this, the time-length measured by an observer who moves relative to the event that is purely time-like in K is always bigger, a phenomenon known as *time-dilation*. In the direction of the movement the lengths of the objects are measured shorter. This phenomenon is known as *length-contraction*, and can be derived immediately from equations (6). When measuring the length of an object from reference frame K' we need to determine the end-points coordinates in the same time moment, so $dt_1' = 0$. For deriving the magnitude of the length-contraction we have to use thus the inverse transformations of (6). We get:

$$dx_1 = \frac{dx_1'}{\sqrt{1 - \frac{u^2}{c^2}}} \rightarrow dx_1' \leq dx_1 \tag{11}$$

The velocity transformations result simply from equations (3),(4) and (8):

$$v_1' = \frac{v_1 + u}{1 + \frac{v_1 u}{c^2}} \quad ; \quad v_1 = \frac{v_1' - u}{1 - \frac{v_1' u}{c^2}} \tag{12}$$

**APPENDIX III. CAUSALITY PRESERVED**

From the Lorentz transformations we already foresee that we have problems if we have reference frames that move relative to each other with a higher velocity than the velocity $c$ of light, since negative numbers under the square root appear. This does not forbid however, to have a special signal that travels with a higher velocity than $c$.

Let us now recall that our belief in the principle of causality was the main reason for which we needed time. It is imperative then that causality should be preserved by the space-time entity we have constructed. The question that makes sense from such a viewpoint is whether causality is preserved within any frame of reference. Let us assume that we have two events *a* and *b*, which are causal: *a* causes *b*. Let us assume that event *a* happens within frame of reference K at spatial location $x_a$ and time $t_a$, and event *b* happens at spatial location $x_b$ at the time moment $t_b$. Let us assume $x_a < x_b$ and causality in frame of reference K means that $t_a < t_b$. This order must be preserved in any inertial frame of reference. If we had an inertial frame of reference where this order is inverted, the principle of relativity would be violated, since the event does not make scientifically sense within this frame of reference. Causality in the frame of reference K means that there should be some signal starting from space point $x_a$ at time moment $t_a$ and arriving at space point $x_b$ in time moment $t_b$. We will consider that this signal is the one that creates a causal connection between the two events, i.e. event b is triggered through this channel. From the $x_a < x_b$ spatial location of the events and from their time moments $t_a < t_b$, one can calculate the propagation speed of this signal. In principle any positive velocity value $v = (x_b - x_a)/(t_b - t_a) > 0$ is acceptable. Let us now use a second K' frame of reference travelling in the direction of the line from point $x_a$ to $x_b$ at a velocity of $u$ (note that this is quite the opposite direction as the one sketched in Figure 1). We learn from the Lorentz transformations, (use of eq. (12) with $u \to -u$) that within this frame of reference:

$$t_b' - t_a' = \frac{t_b - t_a}{\sqrt{1 - \frac{u^2}{c^2}}}\left(1 - \frac{vu}{c^2}\right) \qquad (13)$$

In order to preserve causality ($t_b' > t_a'$) we need thus $uv<c^2$ be satisfied for any $u<c$ value! Since the speed of reference frame K' and the signal from *a* to *b* are not connected, their values can vary independently. We thereby arrive at the conclusion that the value of *v* is also bound by the value of *c*. In other words, in order to preserve the principle of relativity (i.e. to hold causality within any frame of reference) we put forward that there is a speed limit in our Universe, and this is the speed of light! It is not allowable for any signal to be propagated faster than the speed of light, and no frame of reference (i.e. object) can have a speed relative to the other one bigger than the speed of light. These statements must be true in order to preserve the logical consistency of the special theory of relativity.

**APPENDIX IV. THE COSMIC MICROWAVE BACKGROUND RADIATION**
The CMBR is an electromagnetic radiation detectable throughout Universe, and thus measurable on the Earth, too. Arno Penzias and Robert Wilson discovered it by accident in 1964 during radio astronomy and satellite communications experiments. In the signal captured by their specially built horn antenna, the CMBR appeared as a continuously present electromagnetic noise coming from every direction in the Universe. The spectrum of this electromagnetic radiation peaks in the microwave region, and it was found to be characteristic of very low-temperature (around 3K) thermal (or black-body) radiation. This electromagnetic radiation is assumed to be a leftover from the "Big Bang" that created our Universe. Nowadays, using modern apparatus, we are able to map this radiation with great precision at micro Kelvin resolution, and anisotropies from different directions of space are precisely measurable. It was found that apart from randomly distributed, tiny anisotropies, the CMBR measured from Earth has also a prominent "dipole-anisotropy" on milli Kelvin resolution, which results from the motion of Earth (and our group of galaxies) relative to the CMBR's frame of reference [15]. In this way, we have a detectable, special frame of reference, relative to which the dipole anisotropy of the CMBR would vanish. Interestingly, we know this special frame of reference, and it appears to be moving at a speed of around 600 km/s relative to the Milky Way.